\documentclass[5p,twocolumn,times,sort&compress]{elsarticle}

\usepackage{lineno,hyperref}
\usepackage{graphicx}
\usepackage{dcolumn}
\usepackage{bm}
\usepackage{rotating}
\usepackage{longtable}
\usepackage{float}
\usepackage{eucal}
\usepackage{csquotes}
\usepackage{xcolor}

\modulolinenumbers[5]

\begin{document}

\begin{frontmatter}







\title{\centering {Collective enhancement in nuclear level density of $^{72}$Ga and its effect on $^{71}$Ga(n, $\gamma$)$^{72}$Ga capture cross-section}}

\author[label1]{Rajkumar Santra \corref{cor1}}
\cortext[cor1]{Corresponding author at: Tata Institute of Fundamental Research, Mumbai, India.}
\ead{rajkumarsantra2013@gmail.com}
\author[label2]{Balaram Dey}
\author[label3]{Subinit Roy}
\author[label1]{R. Palit}
\author[label1]{Md. S. R. Laskar}
\author[label4]{H. Pai}
\author[label5]{S. Rajbanshi}
\author[label6]{Sajad Ali}
\author[label3]{Saikat Bhattacharjee}
\author[label1]{F. S. Babra}
\author[label3]{Anjali Mukherjee}
\author[label1]{S. Jadhav}
\author[label1]{Balaji S Naidu}
\author[label1]{Abraham T. Vazhappilly}
\author[label1]{Sanjoy Pal}
\address[label1]{Department of Nuclear and Atomic Physics, Tata Institute of Fundamental Research, Mumbai-400005, India.}
\address[label2]{Department of Physics, Bankura University, Bankura-722155, India.}
\address[label3]{Nuclear Physics Division, Saha Institute of Nuclear Physics, Kolkata-700064, India.}
\address[label4]{Extreme Light Infrastructure - Nuclear Physics, Bucharest-Magurele, 077125, Romania.}
\address[label5]{Department of  Physics, Presidency University, Kolkata-700073, India.}
\address[label6]{Government General Degree College at Pedong, Kalimpong, 734311, India.}

\begin{abstract}

The $\gamma$-gated proton spectra measured in the reactions $^{64}$Ni($^{9}$Be, p2n)$^{70}$Ga and $^{64}$Ni($^{9}$Be, pn)$^{71}$Ga, have been utilized to obtain the nuclear level density (NLD)
of $^{71}$Ga and $^{72}$Ga nuclei by using the statistical model (SM) calculations. It is seen that the $\gamma$-gated proton spectrum are reasonably explained by using the large value of the inverse level density parameter ($k$ = 11.2 MeV)
in the NLD prescription of the Fermi gas (FG) model. The large value of $k$ is indicative of the rotational enhancement, which is consistent with the earlier results in other mass regions. Furthermore, a rotational enhancement factor
has been included in the NLD and used in the SM calculation keeping the systematic value of $k$=8.6 MeV and it explains the $\gamma$-gated proton spectrum nicely. The result clearly indicates the presence of collective
enhancement in NLD. Subsequently, the NLD with collective enhancement has been utilized in the TALYS calculation, for the first time, to calculate the $^{71}$Ga(n, $\gamma$)$^{72}$Ga capture cross-section. It is observed that, 
while the FG model without the collective enhancement in the NLD for $^{72}$Ga under predicts the capture data, with the rotational enhancement correction the FG model over predicts the data by similar amount at higher energies. 
However, in the energy range of 0.01 MeV to 0.1 MeV, the FG model corrected for rotational enhancement describes the data quite well. Thus, the present work indicates that collective enhancement, whenever required, should be taken 
into account fro proper description of low energy capture cross section data.



\end{abstract}


\date{\today}
\end{frontmatter}

The basic understanding of a quantum many-body system such as an atomic nucleus is highly crucial to unfold the chemical evolution of the Universe in its present state. 
To perceive the characteristics of the atomic nucleus, usually it is brought under small external perturbations. Nuclear level density (NLD) is one of the important physical 
quantities in order to understand the response of an atomic nucleus to such external perturbations \cite{Bohr, Bethe}. Level density, defined as the number of excited levels 
per unit of excitation energy for a nucleus, plays an important role as one of the key ingredients in the statistical model description of low energy nuclear reactions. 
Consequently, the knowledge of NLD is also important in the modeling stellar nucleosynthesis and evolution \cite{Rauscher}. 
However, NLD provides the highest uncertainties in estimating the reaction cross-section for production of stable and unstable nuclei in the nucleosynthesis process. 
Reliable nuclear level densities are therefore very crucial in the determination of nuclear reaction rates in the field
of nuclear astrophysics \cite{Rauscher1, Rauscher2}. In addition, the enhancement in the NLD due to collective degrees of freedom (such as rotation and vibration) i.e collective
enhancement \cite{Banerjee, Pandit, Pandit1, Mohanto} is another puzzling topic \cite{Pandit1}.
The collective enhancement present in the NLD, especially around the particle separation energy, could have a significant role in the particle capture cross-sections.
Therefore, an accurate extraction of the NLD using a reliable experimental technique is essential to constrain the quantity in estimating the important low energy capture
reaction cross-section as precisely as possible.

Neutron capture on $^{71}$Ga is one of the important astrophysical reactions in the s-process, which not only determines the abundances of gallium isotopes but also affects the abundances of the elements up 
to zirconium \cite{seeger}. Gallium is mostly produced by the weak s-process that takes place in massive stars of more than eight solar masses during convective helium-burning (at temperatures T $\sim$ 30 keV) and 
shell carbon burning (at T $\sim$ 90 keV) \cite{seeger}. A recent simulation indicates that a 50$\%$ change in the neutron capture cross section on one gallium isotope changes the s-abundances of all the following 
isotopes in shell carbon burning by up to 20$\%$ \cite{pig}. In addition, there are three discrepant experimental results on the neutron capture cross-section on $^{71}$Ga at 25 keV \cite{anand, macklin, chaubey}. 

The primary goal of the present work is clearly two fold. First, to investigate the collective enhancement in the NLDs of deformed nuclei $^{72}$Ga and $^{71}$Ga by using the evaporation particle-gamma 
coincidence technique. Secondly, to use the extracted NLD of $^{72}$Ga isotope in the statistical model description of $^{71}$Ga(n,$\gamma$)$^{72}$Ga capture cross sections by upgrading the statistical model 
code TALYS \cite{talys} to include the effect of collective enhancement in the NLD prescription.


Experimentally, the precise measurement of NLD is a very difficult task due to the lack of suitable techniques. The density of low lying excited levels of the nucleus (small in number, well separated, and simple in nature) is generally 
measured directly by counting the levels \cite{RIPL3} and from neutron resonance studies \cite{Huizengaand}. Whereas, the Oslo technique based on particle-gamma coincidence \cite{Schiller}, 
the two-step cascade method \cite{bec}, the $\beta$-Oslo method \cite{spy}, the gamma-ray calorimetry \cite{ull}, etc, are used to  estimate the level density especially up 
to the particle separation energy. Another approach to estimate the level density below as well as above the particle threshold energy is the particle evaporation technique \cite{voi1, dey,rami,rami2,byun}. 
In this technique, the evaporated particles (neutron or proton or alpha) are measured and compared with the statistical model calculations using the different formalisms for level 
density such as  FG (Fermi gas), Gilbert-Cameron, etc. The parameters characterizing the level density of each formalism are varied to explain the evaporated particle spectrum and the study yields the level density of the corresponding 
residual nucleus.The main concern with the particle evaporation technique is to measure the particles from compound nuclear reactions, which can be confirmed by measuring the particles at backward angles and studying the angular distribution of the emitted particles in light-ion induced reactions \cite{voi1, dey,rami,rami2,byun}.
Very recently, we adopted a new approach to select the evaporated particles only from compound nuclear 
reactions by using the gamma-gated particle spectra \cite{Santra}. The same approach, as discussed in Ref. \cite{Santra} has been used in the present work.

\begin{figure}
\begin{center}
\includegraphics[height=4.0cm, width=8.4cm]{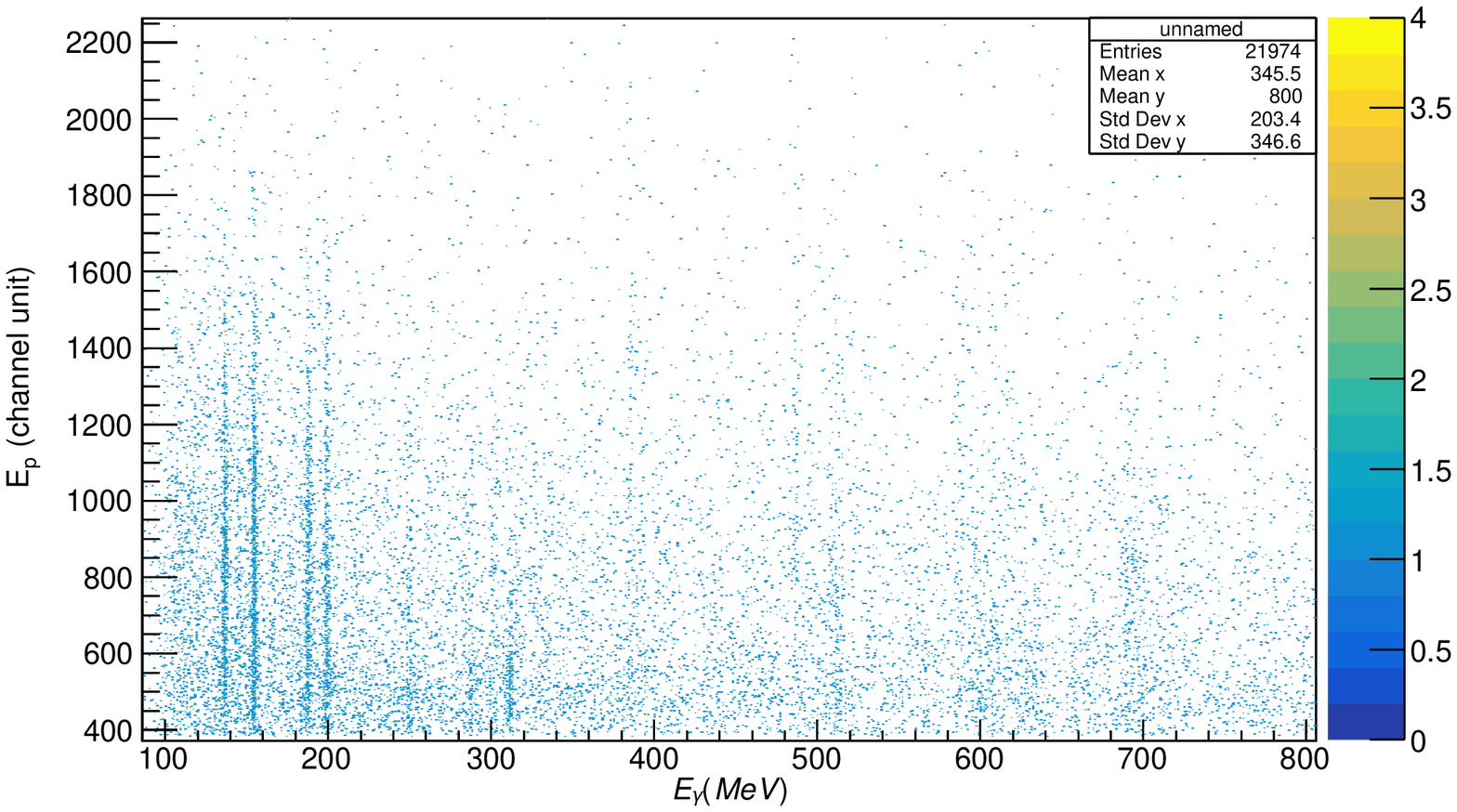}
\includegraphics[height=4.0cm, width=8.cm]{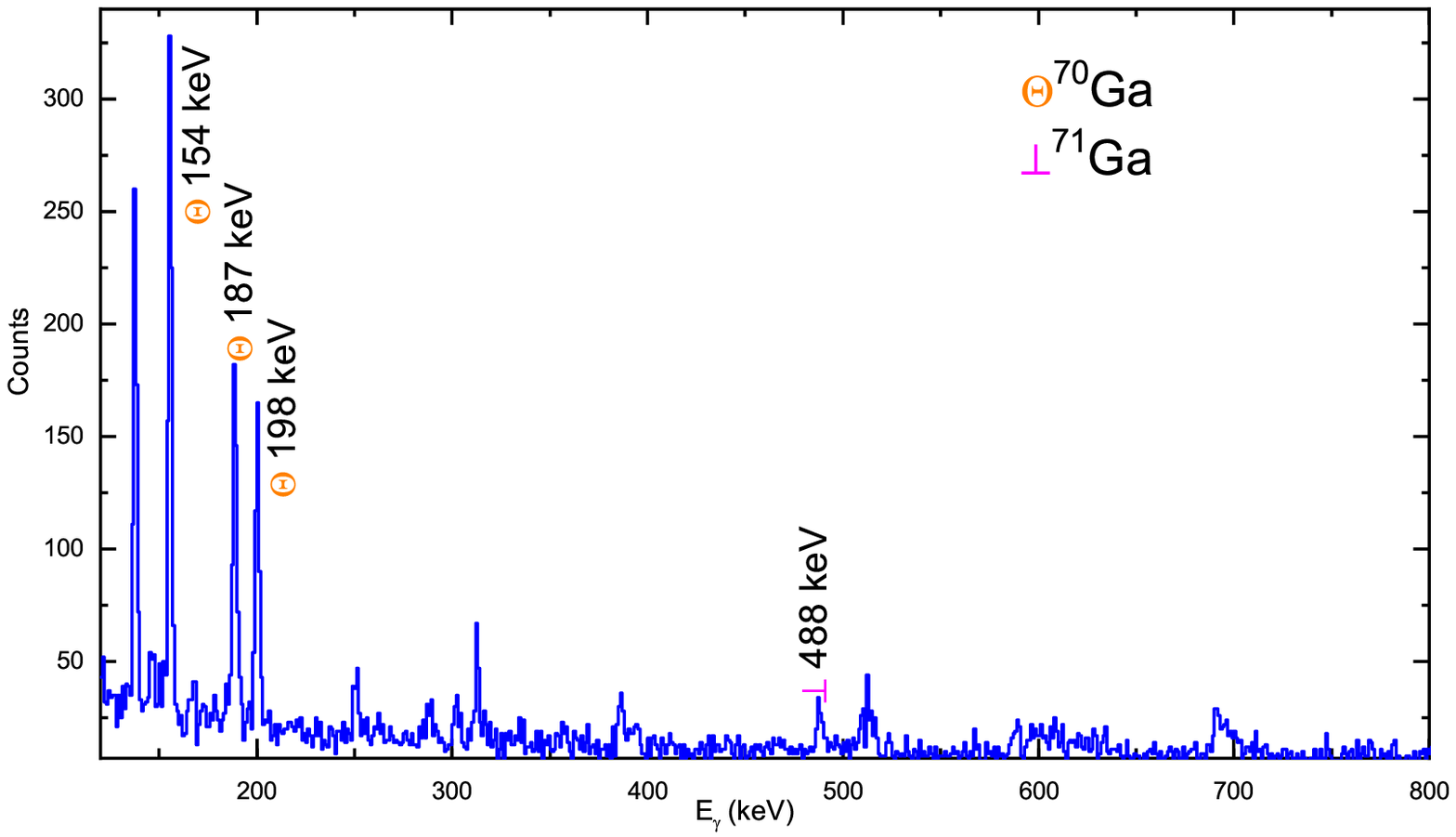}
\caption{\label{fig12} (Color online) [Top panel] p-$\gamma$ coincidence matrix extracted from raw particle-$\gamma$ matrix. [Botom panel] Projected gamma energy spectrum. 
Symbols with $\gamma$-energy indicate $\gamma$-lines of different residual nuclei associated with proton emitting channels.}
\end{center}
\end{figure}

In this work, we have measured the $\gamma$-gated evaporated protons coming from the reactions $^{64}$Ni($^{9}$Be, p2n$\gamma$)$^{70}$Ga and $^{64}$Ni($^{9}$Be, pn$\gamma$)$^{71}$Ga. 
Particle-gamma coincidence has been performed to eliminate the events from target contamination and pre-equilibrium emission. The $\gamma$-gated proton spectra have been utilized 
to obtain the level densities of residual nuclei. The obtained level density for $^{72}$Ga has further been used to calculate the $^{71}$Ga(n,$\gamma$) capture reaction cross-section 
as a function of neutron energy.




\begin{figure}
\begin{center}
\includegraphics[height=10.5cm, width=9.0cm]{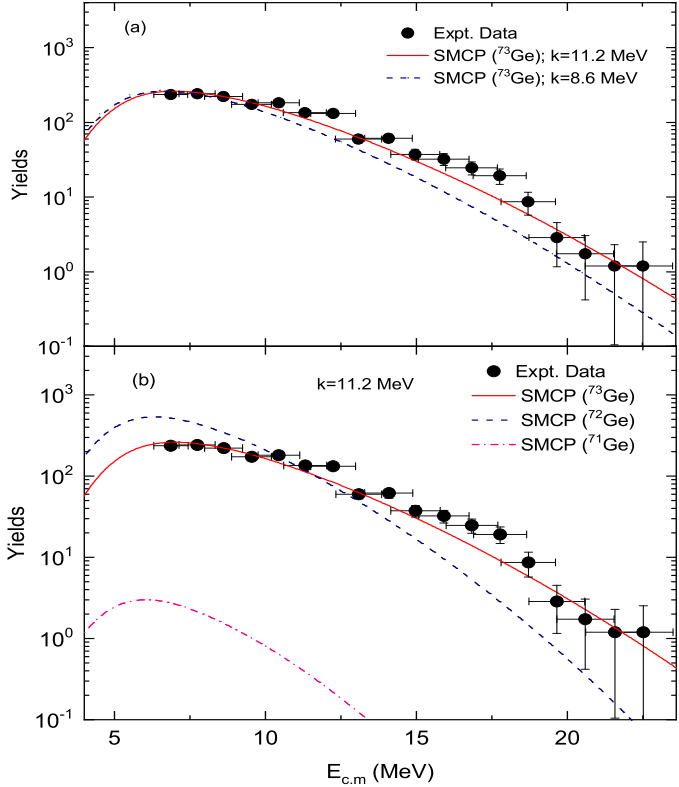}
\caption{\label{fig3} (Color Online) Filled symbols represent the experimental $\gamma$-gated proton spectrum. Lines represent the statistical model calculated proton (SMCP) spectrum. [Top panel]
Red solid and blue dashed lines represent SMCP spectrum considering 1$^{st}$ chance proton emission from CN, with k= 11.2 MeV and k=8.6 MeV, respectively. 
[Botom panel] Red continuous, blue dashed and pink dash-dotted lines represent SMCP spectrum considering 1$^{st}$ , 2$^{nd}$ and 3$^{rd}$ chance proton emission from CN, respectively, with k=11.2 MeV.}
\end{center}
\end{figure}

The $^{9}$Be beam of 30 MeV energy with an average current of 5 nA from BARC-TIFR Pelletron Linac Facility, Mumbai, was used to bombard a self-supporting $^{64}$Ni target of thickness 500 $\mu$g/cm$^2$, populating the $^{73}$Ge compound nucleus (CN). 
The outgoing protons were detected by using CsI(Tl) detectors with the angle of coverage from 22$^\circ$ to 67$^\circ$. The 14 Compton-suppressed Clover detectors were used to detect the de-exiting discrete $\gamma$-rays coming from the 
residual nuclei. XIA-LLC-based digital data acquisition system was used to store the data in list mode. Tantalum foil of thickness 30 mg/cm$^2$ was used to stop the elastically scattered events from entering the detectors. The proton spectrum is calibrated using
$^{229}$Th source with pulse height defect correction based on light output vs. energy data of CsI(Tl) in Ref.\cite{dell}.  
Multi parameter time stamped based coincidence search program (MARCOS) \cite{palit} was used to construct the p-$\gamma$ matrix as shown in Fig. \ref{fig12}. From this matrix, proton yield spectra were extracted gated by 
characteristic 154.9 (5$^-$ $\rightarrow$ 4$^-$), 187.6 (4$^-$ $\rightarrow$ 2$^-$), 198 keV $\gamma$-rays from the residual $^{70}$Ga nucleus and 488 keV $\gamma$-ray from $^{71}$Ga nucleus \cite{ARNELL}. 
Finally, the $\gamma$-gated proton spectra are added after efficiency correction 
( Y$_{Total}$= $\frac{Y_{154} }{ \epsilon_{154}}$ + $\frac{Y_{187} }{ \epsilon_{187}}$ + $\frac{Y_{198} }{ \epsilon_{198}}$ + $\frac{Y_{488} }{ \epsilon_{488}}$ ) 
to visualize the shape of the final proton spectrum with higher statistics as shown in Fig. \ref{fig3}. The quantities, $\epsilon_i$ and Y$_i$, are the detection efficiency of i$^{th}$ $\gamma$-ray
and the yield of proton that are in coincidence with the $i^{th}$ $\gamma$-ray.



\begin{figure}[t]
\centering
\setlength{\unitlength}{0.05\textwidth}
\begin{picture}(10,13.0)
\put(-0.5,-0.80){\includegraphics[width=0.5\textwidth,height=0.7\textwidth, angle = 0]{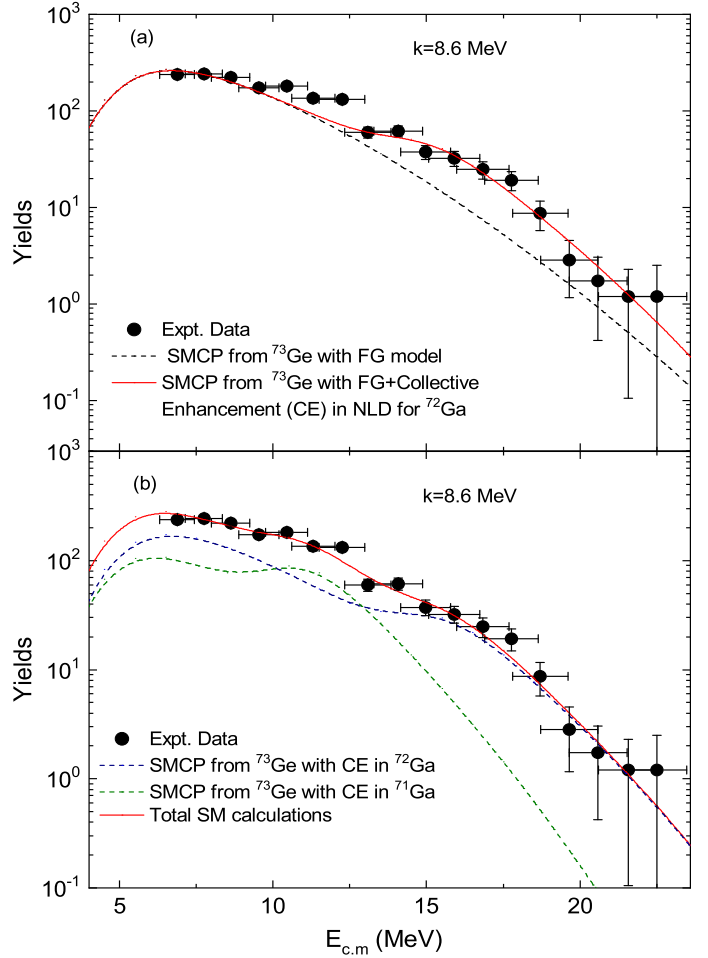}}
\end{picture}
\caption{\label{fig4}(Color online) Filled symbols represent the $\gamma$-gated proton spectrum. Lines represent the statistical model calculated proton (SMCP) spectrum. 
[Top panel] Red continuous line represents SMCP with k=8.6 MeV considering 1$^{st}$ chance proton emission from CN $^{73}$Ge and including collective enhancement factor in NLD of $^{72}$Ga residual nucleus.
[Botom panel] Purple dashed line represents SMCP with k=8.6 MeV considering 1$^{st}$ chance proton emission from CN $^{73}$Ge and including collective enhancement factor in NLD of $^{72}$Ga
residual nucleus. Pink dashed line represents SMCP with k=8.6 MeV considering 2$^{nd}$ chance proton emission from CN $^{73}$Ge and including collective enhancement factor in NLD of $^{71}$Ga residual nucleus. 
Red continuous line represents the total spectrum with weitage factor as Y$_{total}$ =Y(1$^{st}$ chance) + 0.3$\times$Y(2$^{nd}$ chance).}
\end{figure}

In order to investigate the NLD of residual nuclei, the $\gamma$-gated proton spectrum has been compared with the statistical model calculations using the code CASCADE \cite{cas} with the Fermi-Gas (FG) model of NLD given by
\begin{equation}
\rho(E^*,J) = \frac{2J + 1}{12\theta^{3/2}} \sqrt{a} \frac{exp(2\sqrt{a U})}{U^{2}}
\end{equation}
where, U$=$E$^*$$-$$\frac{J(J+1)}{\theta}$$-$S$_p$$-$$\Delta$P, E$^*$ and J are the initial excitation energy and the angular momentum of the nucleus, respectively, $\theta=\frac{2I_{eff}}{h^2}$, with I$_{eff}$, S$_p$ and $\Delta$P 
being the effective rigid-body moment of inertia, the proton separation energy and the pairing energy respectively. The quantity $a$ in Eq.1 is the level density parameter.
Ignatyuk prescription \cite{igna2} of level density parameter $a$, which takes into account the shell effects as a function of excitation energy is adopted and it is expressed as 
\begin{equation}
a = \tilde{a}[1 + \frac{\delta S}{U} [1-\exp(-\gamma U)]] 
\end{equation}
where, asymptotic level density parameter $\tilde{a}$ = A/$k$ and $k$ is the inverse density parameter. $\delta S$ is the ground-state shell correction defined as the difference of the experimental and theoretical (liquid drop) masses. 
The inverse of $\gamma$ in the exponent of Eq.2, given by $\gamma^{-1}$ = $\frac{0.4A^{4/3}}{\tilde{a}}$ is the rate at which the shell effect is damped with the increase in excitation energy. 
The optical model potential parameters for proton transmission coefficient are taken from Refs. \cite{OPM}. The moment of inertia of the CN is taken as $I_{eff} = I_0(1 + \delta_1 J^2 + \delta_2 J^4$), where $I_0$(= $\frac{2}{5}MA^{5/3}r_0^2$) is the moment 
of inertia of a spherical nucleus, $\delta_1$ (= 1.0 $\times$ 10$^{-4}$) and $\delta_2$ (= 2.0 $\times$ 10$^{-6}$) are the deformability parameters, r$_0$=1.25 fm is the radius parameter and $J$ is 
the total spin of the nucleus. It is seen that the choice of proton optical potential has negligible effect on the slope of the spectrum. The deformability parameters have also negligible effect on 
slope ($\delta_1$ change from 10$^{-6}$ to 10$^{-4}$ and $\delta_2$ change from 10$^{-8}$ to 10$^{-6}$). The slope of the spectrum mainly depends on the inverse level density parameter $k$, which 
is generally varied to explain the evaporated particle spectrum by using $\chi^2$ minimization technique. 

The $\gamma$-gated proton spectrum along with the statistical model calculations is shown in Fig. \ref{fig3}. It is observed that the systematic value of $k$ (8.6 MeV) is unable to explain the experimental 
data above 10 MeV as shown in Fig. \ref{fig3}(a). Indeed, a large value of $k$ (11.2 MeV) is required to explain the experimental data considering the emission of first chance protons 
from CN as shown in Fig. \ref{fig3}(a). This observation is consistent with the previous results obtained in refs. \cite{Banerjee, Pandit}.
The Ground-state quadrupole deformations of $^{71}$Ga and $^{72}$Ga are -0.207 and -0.215, respectively as have been reported by P. Moller et. al \cite{moeller}. Moreover, rotational bands have also been observed up to the 4.2 MeV excitation energy in $^{71}$Ga \cite{Stefanescu}. One can expect also a rotational band structure at high spin in $^{72}$Ga.Therefore, a rotational enhancement in the Gallium isotopes is in agreement with the structural study. So the collective enhancement due to ground state deformation of residual nucleus could be the possible reason for this large $k$ value. 
On the other hand, as the evaporated proton spectrum is gated with the $\gamma$-rays of $^{70}$Ga and $^{71}$Ga residual nuclei, it could also contain the contributions of protons coming out after one or two neutrons (1n or 2n) 
are emitted from the CN. Therefore, a statistical model calculation has been carried out by using $k$=11.2 MeV to obtain the proton spectrum after 1n and 2n emissions and compared with the experimental data 
as shown in Fig.\ref{fig3}(b). It should be mentioned that the same normalization has been used in each case (protons from the first chance, protons after 1n emission, and protons after 2n emission) while 
comparing with the experimental data. It is seen that there could have been some contributions of protons after 1n emission. However, the proton after 2n emission is negligibly small as shown in Fig. \ref{fig3}(b). 

\begin{figure}[t]
\centering
\setlength{\unitlength}{0.05\textwidth}
\begin{picture}(10,11.5)
\put(-0.1,-1.00){\includegraphics[width=0.49\textwidth,height=0.62\textwidth, angle = 0]{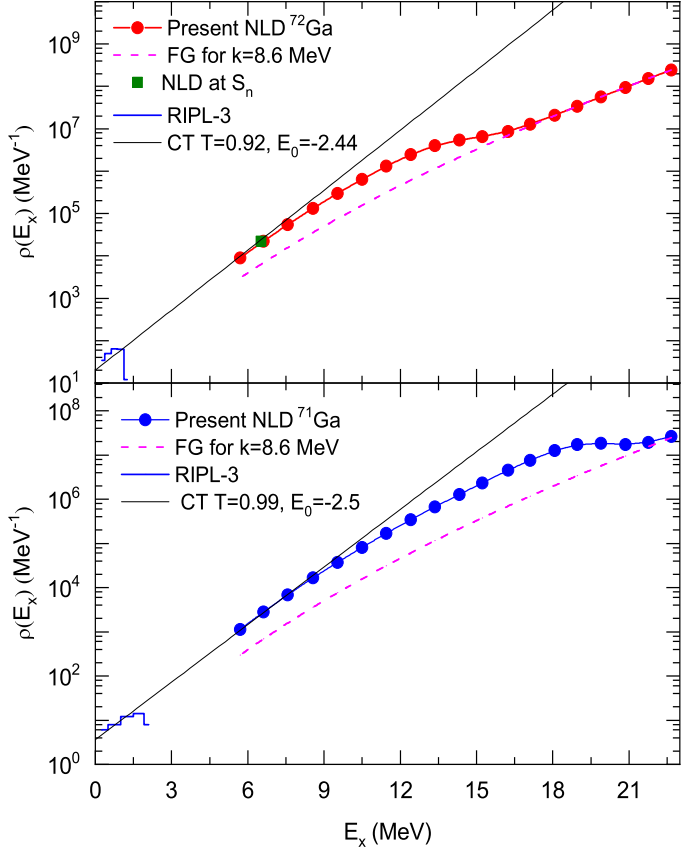}}
\end{picture}
\caption{\label{fig5} (Color online) The enhanced Nuclear level densities as a function of excitation energy are shown as used in the CASCADE. Histogram represents the cumulative level distribution computed from the levels} given in RIPL-3 \cite{RIPL3}
\end{figure}

A large value of $k$ = 11.2 MeV from the previous statistical model analysis with FG model for NLD and the bumps around the centre of mass energies of 12 MeV and 15 MeV, could be a possible reason for the existence of collective enhancement in NLD due to the deformation of the residual nuclei. 
In a subsequent analysis, the $\gamma$-gated proton spectrum has been compared with the statistical model prediction of proton (SMCP) spectrum by including collective enhancement factor in NLD of $^{72}$Ga residual nucleus as shown in Fig.\ref{fig4}.
The model calcualtion has been carried out by considering 1$^{st}$ chance proton emission with the systematic value of $k$=8.6 MeV (which best explains the low energy part of the proton spectra below 10 MeV) and 
a rotational enhancement factor ($K_{rot}$) has been inlcuded as $\rho$=$\rho_{int} K_{rot}$, where $\rho_{int}$ is intrisic level density.  The factor K$_{rot}$ is introduced with the intrinsic single particle level density to consider the contribution of rotational enhancement factor in the NLD. Microscopic shell model studies \cite{Banerjee} have predicted that 
for nuclei with finite ground state deformation, rotational collectivity causes large enhancement of NLD (K$_{rot} \approx$ 100). With increasing excitation energy, the deformation effect gradually dies down due to the intrinsic motion. 
The deformed shape of the residual nucleus changes into spherical shape and rotational levels die out. Due to this shape transition, fade out of collective enhancement in NLD is observed after a certain excitation energy.
The energy dependent Fermi function has been used in the rotational enhancement factor given as 
\begin{equation}
K_{rot}= (\sigma^2_\perp-1)\times \it{f(E)} + 1
\end{equation}
\begin{equation}
f(E)= \frac{1}{1+ \exp(\frac{E-E_{cr}}{d_{cr}})}
\end{equation}
The spin cut off parameter, $\sigma_\perp^2$ is replaced by $\lambda$T, where T [=$\sqrt(U/a)$] is nuclear temperature and $\lambda$ is treated as the magnitude of the enhancement parameter. The parameters $\lambda$, E$_{cr}$ and d$_{cr}$ are generally 
varied to explain the experimental data. It is observed that the the experimental data below 10 MeV and above 13 MeV are nicely described by the result of the statistical model calcualtion considering the first chance protons and 
including the rotational enhancement in NLD of $^{72}$Ga, as shown in Fig. \ref{fig4}(a). However, the experimental data in between 10 to 13 MeV centre of mass energy cannot 
be explained. The conjecture is the presence of protons from 2$^{nd}$ chance emission in that energy domain. Therefore, the proton spectrum has also been calculated considering the level density and including the rotational enhancement 
factor of $^{71}$Ga residual nucleus and shown in Fig. \ref{fig4}(b). Finally, the 1$^{st}$ and 2$^{nd}$ chance statistical model calculated proton (SMCP) spectra are added with weightage obtaining the total yield as 
Y$_{total}$= Y(1$^{st}$ chance) + $w$$\times$ Y(2$^{nd}$ chance), which nicely explains the $\gamma$-gated proton spectrum as shown in Fig. \ref{fig4}(b). A $\chi^2$ minimization procedure, with parameters from two rotational enhancement factors 
corresponding to the NLDs of residual nuclei $^{72}$Ga and $^{71}$Ga and the weight factor $w$ in the total yield, has been carried out to fit the experimental spectrum. The best fit parameters 
of the rotational enhancement factor in NLD are $\lambda$=4.0 $\pm$ 0.5, E$_{cr}$=14 $\pm$ 0.6 MeV and d$_{cr}$=0.8 $\pm$ 0.1 MeV for $^{72}$Ga, and $\lambda$=5.0 $\pm$ 0.5, E$_{cr}$=19 $\pm$ 0.7 MeV and d$_{cr}$=0.7 $\pm$ 0.12 MeV for $^{71}$Ga 
and the weight factor $w$=0.3. 
The fit clearly indicates that the collective enhancement in NLD due to the deformation of residual nucleaus is required to explain the experimental data. The nuclear level densities of the residual nuclei $^{72}$Ga and $^{71}$Ga including the collective enhancements 
used in the statistical model calculation for explaining the $\gamma$-gated spectrum are shown in Fig. \ref{fig5}. The NLD of $^{72}$Ga is normalized with the experimental NLD at S$_n$ obtained from neutron resonance spacing \cite{NLD_Bn}. 
 
In Fig. 4, two different model nuclear level densities have been plotted to compare with the present level densities of $^{72}$Ga and $^{71}$Ga and also compared with the cumulative level distribution computed from the levels given in the RIPL-3 \cite{RIPL3} at low excitation energy. The constant temperature (CT) model, which is known to best describe the level densities in the low excitation energy domain up to the particle 
threshold energy, is represented by solid curve. Also, shown the prediction of FG model with the systematic value of the parameter $k$ = 8.6 MeV and not including the collective enhancement contribution (dashed line).
The CT model level density is given as $\rho_{\rm CT}$ = $\frac{1}{T}$e$^{(E-E_0)/T}$, where T and E$_0$ are free parameters. It should be mentioned that the systematic values of CT model parameters 
(E$_0$=-2.44 MeV, T=0.92 MeV), obtained from the fitting of the NLD data from RIPL-3 and the data at the neutron separation energy (S$_n$ \cite{NLD_Bn} are used while comparing with the level density of $^{72}$Ga from the present work. As the experimental NLD of $^{71}$Ga at S$_n$ (from neutron resonance studies) is not available, 
the present NLD of $^{71}$Ga is normalized with the results obtained from CT model by using the value of E$_0$=-2.5 MeV, T=0.99 MeV based on RIPL-3 data. The important aspects of the comparisons in Fig. \ref{fig5} should be emphasized here. 
It is seen that the extracted density that includes the enhancement effect follows the trend of the CT model prediction locally around the normalization point at S$_n$. However, with  increasing excitation energy beyond the neutron separation 
energy, as the collective enhancement effect wanes, the NLD from the present work deviates from CT model prediction and gradually coincides with the prediction of FG Model without enhancement using $k$=8.6 MeV. 
Thus, the excitation energy dependence of the extracted NLD, $\it {i.e.}$, matching the CT prediction at lower excitations and that of the FG model prediction without the deformation effect at higher excitations, 
indicates that the consideration of rotational enhancement in NLD can play a significant role in the low energy neutron capture process.


\begin{figure}[t]
\centering
\setlength{\unitlength}{0.05\textwidth}
\begin{picture}(10,7.0)
\put(-0.4,-1.00){\includegraphics[width=0.5\textwidth,height=0.4\textwidth, angle = 0]{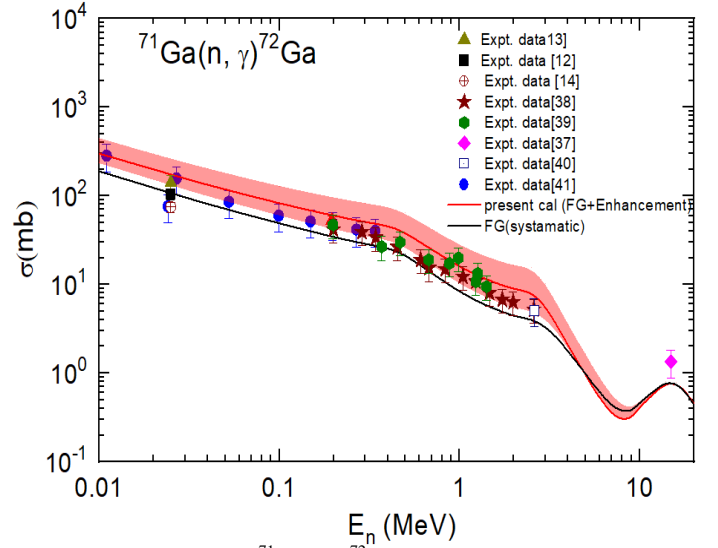}}
\end{picture}
\caption{\label{fig6} (Color online) $^{71}$Ga(n, $\gamma$)$^{72}$Ga capture cross-section as function of neutron energy. 
Reported experimental data are compared with the results obtained from TALYS calculation using the present enhanced level density parameter.}
\end{figure}

Furthermore, the collective enhancement in NLD has been included in TALYS 1.95 code \cite{talys}, for the first time, to investigate the effect of collective enhancement in NLD on neutron capture cross-section. 
A rotational enhancement function for $^{72}$Ga nucleus has been included in the FG model prescription of NLD  $\rho(E_x)$=$\rho_{int}(E_x) K_{rot}$  in TALYS 1.95 code to estimate the $^{71}$Ga(n, $\gamma$)$^{72}$Ga 
capture cross-section by using the systematic value of inverse level density parameter (k = 8.6 MeV). 
In the TALYS code, default global neutron optical model potential has been used for $^{71}$Ga+n and a generalized Lorentzian form \cite{kopecky} of E1 $\gamma$-ray strength function has been considered for $^{72}$Ga. It is found that (Fig. \ref{fig6}), the NLD without 
the collective enhancement underpredicts the neutron capture cross-section data, while the inclusion of collective enhancement in NLD can explain the data much better, especially in the low excitation energy domain. The lower and upper limit (red dashed line in Fig. \ref{fig6})
of the present cross-section arises due to the uncertainties of parameters characterizing the collective enhancement function as discussed earlier. As can be seen from Fig. \ref{fig6}, the neutron capture cross-section 
with including collective enhancement is $\sim$2.1 times greater than that of FG model with systematic LD parameter in the energy range of 25 - 90 keV and matches the data quite well in the low energy region. At higher energies the 
capture data, however, fall within the lower limit of the curve including the enhancement effect. It should be mentioned here that there exists some  difference in the relative normalization between the 
different capture data sets. If the normalization is properly taken care of, the matching with the present calculation will improve further. 
The present result indicates that with the inclusion of collective enhancement in the NLD, the calculated $^{71}$Ga(n, $\gamma$)$^{72}$Ga cross-sections are closer to the direct measurement data.
In the low excitation energy domain, there are three discrepant results for $^{71}$Ga(n, $\gamma$)$^{72}$Ga cross-sections at 25 keV, which are 104 $\pm$ 14 mb \cite{anand}, 
140 $\pm$ 30 mb \cite{macklin} and 75 $\pm$ 10 mb \cite{chaubey}. Interestingly, the result at 25 keV from the present work is consistent with the result obtained in Ref. \cite{macklin}.
It is worthwhile to mention that overall in the low energy region, which is of astrophysical importance, the NLD with the collective enhancement that explains 
the proton evaporation spectrum, better describes the available neutron capture data. It should be added that the $\gamma$-ray strength function could also play an 
important role to estimate the neutron capture cross-section. 
Therefore, level density and $\gamma$-ray strength function are both essential to estimate the astrophysical reaction cross-section as precisely as possible.

In summary, the measured $\gamma$-gated proton spectrum is predominantly from the compound nuclear process, which is ensured by the coincidence measurement with selected $\gamma$ rays from the residues of p2n and pn channels following 
the $^9$Be+$^{64}$Ni reaction. The $\gamma$ gated proton spectrum has been utilized to obtain the level densities of residual nuclei ($^{72}$Ga and $^{71}$Ga). It is observed that collective enhancement
factors due to the deformation of residual nuclei are required to be included in the NLD prescription of the Fermi-Gas (FG) model to explain the experimental $\gamma$-gated proton spectrum. Furthermore, the obtained level 
density for $^{72}$Ga has been used in the TALYS reaction code to calculate the $^{71}$Ga(n, $\gamma$) $^{72}$Ga capture cross-section. It is seen that NLD with collective enhancement explains the neutron capture data better than the 
 NLD without collective enhancement. The present results indicate that collective enhancement in NLD could play an important role in order to understand the production of nuclei through the neutron capture process 
precisely. Therefore, a systematic and exclusive investigation is required in near future to understand the effect of collective enhancement on neutron capture cross-section to provide a reliable prediction of 
astrophysical abundances.



\begin{center}
$\textbf{Acknowledgements}$
\end{center}
We would like to thank Dr. D. Pandit, VECC for his help and advice in this project. Authors thank the BARC-TIFR PLF staff for uninterrupted, steady  beam during the experiment.This 
work is supported by the Department of Atomic Energy, Government of India (Project Identification No. RTI 4002), and the Department of Science and Technology, 
Government of India (Grant No.IR/S2/PF-03/2003-II).

\end{document}